\documentclass[%
 aps,
 prb,
 twocolumn,
 reprint,%
citeautoscript,
showkeys
]{revtex4-1}

\usepackage{chemformula}
\usepackage{graphicx}%
\usepackage{dcolumn}%
\usepackage{siunitx}
\usepackage{bm}%
\usepackage{placeins}
\begin{document}

\title[]{Drug design principles from electric field calculations: understanding SARS-CoV-2 main protease interaction with X77 non-covalent inhibitor.} %

\author{V. Vaissier Welborn}
 \email{vwelborn@vt.edu}
\affiliation{Department of Chemistry, Virginia Tech, Blacksburg, VA 24061%\\This line break forced with \textbackslash\textbackslash
}%

\date{\today}%

\begin{abstract}
Fast and effective drug discovery processes rely on rational drug design to circumvent the tedious and expensive trial and error approach. However, accurate predictions of new remedies, which are often enzyme inhibitors, require a clear understanding of the nature and function of the key players governing the interaction between the drug candidate and its target. Here, we propose to calculate electric fields to explicitly link structure to function in molecular dynamics simulations, a method that can easily be integrated within the rational drug discovery workflow. By projecting the electric fields onto specific bonds, we can identify the system components that are at the origin of stabilizing intermolecular interactions (covalent and non-covalent) in the active site. This helps to significantly narrow the exploration space when predicting new inhibitors. To illustrate this method, we characterize the binding of the non-covalent inhibitor X77 to the main protease of SARS-CoV-2, a particularly time-sensitive drug discovery problem. With electric field calculations, we were able to identify 3 key residues (Asn-142, Met-165 and Glu-166), that have functional consequences on X77. This contrasts with the nearly 20 residues reported in previous studies as being in close contact with inhibitors in the active site of the protease. As a result, the search for new non-covalent inhibitors can now be accelerated by techniques that look to optimize the interaction between candidate molecules and these residues.
\end{abstract}

\maketitle

\section{\label{introduction}Introduction}
Enzymes are the target of many drug molecules, which bind the active site in place of the natural substrate, preventing the reaction to take place.\cite{supuran2017advances,vassar2014bace1,baillie2016targeted,stout2014exogenous} These drug molecules, or enzyme inhibitors, are traditionally developed over a tedious trial and error process that drives up the cost and duration of drug discovery.\cite{clout2019drug,nass2018accelerating} Although recent advances in combinatorial chemistry\cite{huang2016protein,miles2018peptide,liu2017combinatorial,frei2019dynamic} and high-throughput screening\cite{blay2020high,lim2018microfluidic,hall2016fluorescence,jordi2018high} make it possible to synthesize and test a large number of compounds in record time, a more rational approach is needed.\cite{van2016pharmacotherapy,cozza2017development} 

Rational drug design is based on the prediction, often from computational approaches, of a few candidate molecules to be tested for potency.\cite{jarvis2019essential,ramirez2016computational,do2018steered} The accuracy of the prediction depends on our knowledge of the target enzyme as well as the method used. Molecular docking approaches are the most common in drug design, scoring candidate molecules on their ability to bind the active site.\cite{torres2019key,de2016molecular,nakano2018computational} This scoring is usually made on single snapshots, optimizing close contacts between the enzyme and the candidate molecule. Molecular dynamics (MD) is sometimes used to refine molecular docking predictions, providing conformational ensembles from which we can estimate binding energies.\cite{sledz2018protein,do2018steered,hernandez2016current,rahman2020virtual} More recently, machine learning algorithms were trained on such data and helped produce families of compounds mimicking one or more known inhibitors.\cite{zhang2017machine,schneider2019mind,yang2018silico} Although informative, these studies focus on structural information and often lack functional information.

In this paper, we propose to to directly link structure to function via electric field calculations in an ensemble that characterize picosecond to microsecond timescales. Classical in nature, the magnitude and direction of electric fields can be calculated from MD, providing an accurate representation of electrostatics in the underlying force field.\cite{welborn2019fluctuations} This provides a unique handle on reactivity because the electric fields indicates the preferred direction of electron motion in a model where electrons are not explicitly represented. Further, we can decompose electric fields into contributions from each system component (protein residues, solvent molecules, etc.), which allows the identification of the key players in the system.\cite{nash2020electric} Such electric field calculations were proven valuable to understand the molecular origin of enzymatic activity ($k_\text{cat})$ and predict functional mutations in synthetic enzymes.\cite{welborn2018computational,vaissier2018computational,welborn2019fluctuations} In this case, electric fields were projected onto the bonds that break or form during the reaction to quantify the role of the greater environment on catalysis. However, electric field calculations have a broader range of applications. Indeed, the binding of an inhibitor in the active site of an enzyme will result in a series of local and global conformational changes that will in turn affect the charge distribution. As a result, electric fields can be used in rational drug design to identify the system components governing enzyme-inhibitor interactions. This is particularly useful to narrow the search space in cases where many residues could be involved in substrate binding. The main protease of SARS-CoV-2, for example, has a large and open active site where as many as 20 residues are in close contact with the substrate.\cite{jeong2020therapeutic,ullrich2020sars}

\begin{figure}
\includegraphics[width=0.5\textwidth]{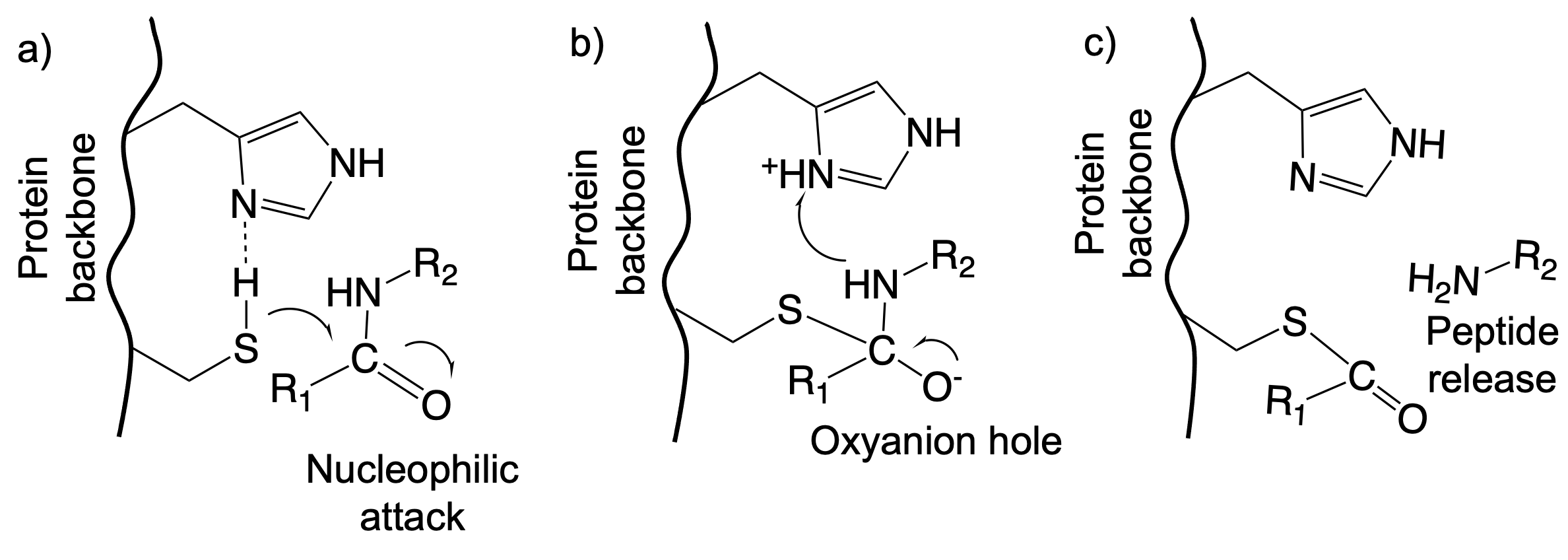}% Here is how to import EPS art
\caption{\label{figure1}Catalytic mechanism of the cysteine protease in SARS-CoV-2. Cys-145 initiates bond cleavage with a nucleophilic attack. The intermediate is stabilized by the formation of an oxyanion hole, supported by His-41. The release of the first peptide regenerates His-41 before Cys-145. }
\end{figure}

SARS-CoV-2, the RNA virus responsible for Covid-19 illustrates perfectly the need for a faster route to drug discovery. There is a real urgency in finding a remedy to the disease that has paralyzed society since the first outbreak in December 2019.\cite{estrada2020topological,jeong2020therapeutic} SARS-CoV-2 main protease, M$^\text{pro}$ or 3CL$^\text{pro}$, catalyzes the cleavage of peptide bonds during viral replication, which makes it a great target for drug development.\cite{ngo2020computational,jeong2020therapeutic,ullrich2020sars} M$^\text{pro}$ initiates peptide bond cleavage with the nucleophilic attack of a cysteine residue (Cys-145) in cooperation with a histidine residue (His-41), forming a catalytic dyad (see Figure \ref{figure1}). 
Strategies to rapidly roll-out a Covid-19 treatment involve repurposing existing drugs and generating compounds that mimic known inhibitors to other viral proteases.\cite{estrada2020topological,dai2020structure,baby2020targeting,ullrich2020sars} For example, aldehyde inhibitors that covalently bind Cys-145 were proven particularly potent for cysteine proteases, which inspired some Covid-19 studies.\cite{zhang2020crystal} Others started to look at the development of potentially less toxic non-covalent inhibitors. The combination of covalent, hydrogen and van der Waals interaction in M$^\text{pro}$ active site involves many residues, namely His-41, Ser-46, Met-49, Tyr-56, Phe-140, Leu-141, Asn-142, Ser-144, Cys-145, His-163, His-164, Met-165, Glu-166, Leu-167, His-172, Phe-185, Asp-187, Gln-189, Thr-190 and Gln-192.\cite{jeong2020therapeutic,ullrich2020sars} Without further knowledge, drug candidates are selected on the basis of structural information such as number of contacts, binding energy, dissociation constant, etc.\cite{dai2020structure,ngo2020computational} Some have tried to circumvent this issue by using machine learning approaches\cite{zhavoronkov2020potential} but no potent inhibitor has been reported so far. A more efficient, long-term, solution would be to produce an inhibitor that specifically targets M$^\text{pro}$, which requires a deeper understanding of the inhibitor-M$^\text{pro}$ interaction.\cite{macchiagodena2020inhibition,jeong2020therapeutic} 
 
\begin{figure}
\includegraphics[width=0.3\textwidth]{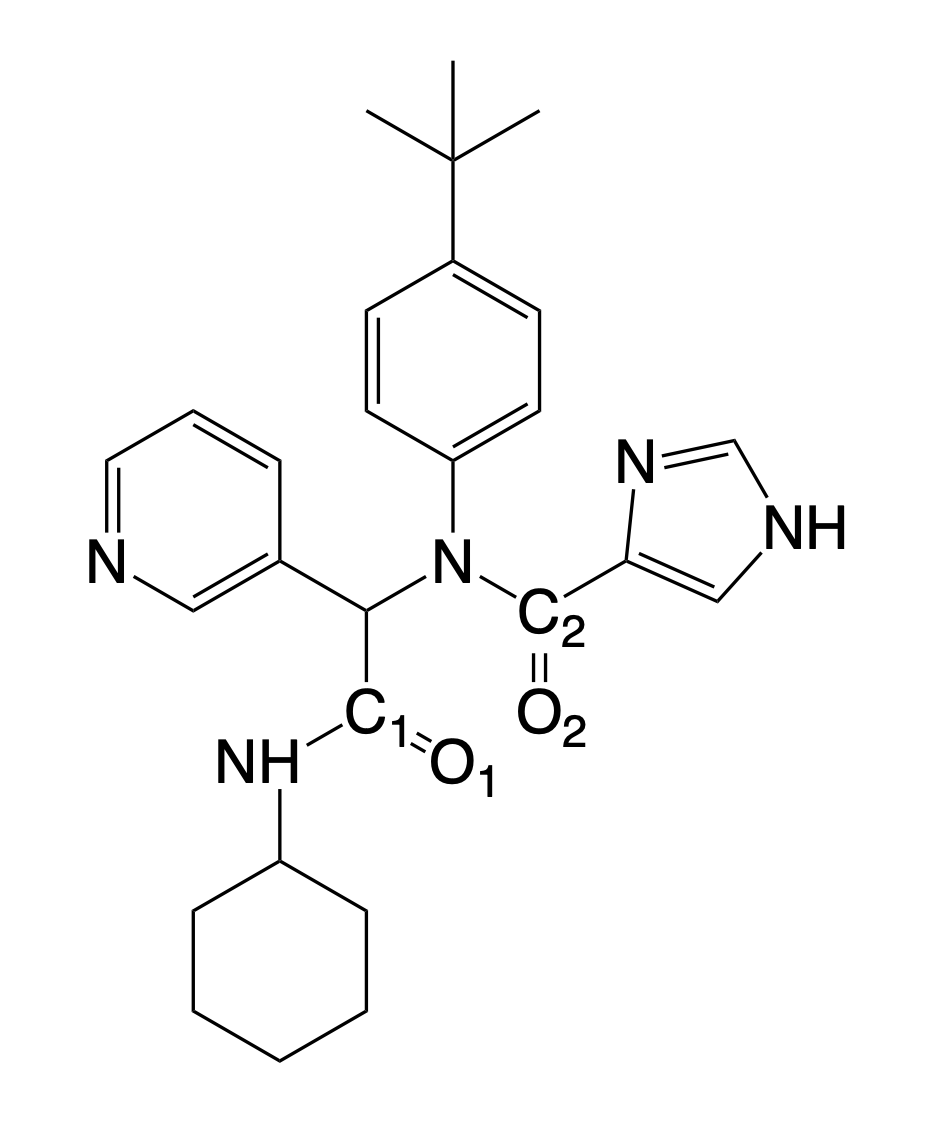}% Here is how to import EPS art
\caption{\label{X77}Chemical structure of the non-covalent inhibitor X77 or N-(4-tert-butylphenyl)-N-[(1R)-2-(cyclohexylamine)-2-oxo-1-(pyridin-3-yl)ethyl]-1H-imidazole-4-carboxamide. The two C=O bond of the amid groups are labelled C$_1$=O$_1$ and C$_2$=O$_2$ for convenience.}
\end{figure}

In this paper, we analyze the interaction of the inhibitor X77 (Figure \ref{X77}) with M$^\text{pro}$ to derive Covid-19 drug design principles. X77 has been reported in many studies as a promising non-covalent inhibitor and has served as the basis of mimetic studies to get novel antiviral agents.\cite{zhavoronkov2020potential, andrianov2020computational} Our work will facilitate the design of X77-like compounds for the inhibition of  M$^\text{pro}$ by linking structural information to function over multiple time scales. The results presented here allowed us to identify specific residues that interact with X77 from atomistic simulations.

\section{\label{methods}Computational Methods}
The atomistic simulations presented in this paper were performed using the PDB structure of SARS-CoV-2 main protease complexed with X77 as a starting point (PDB ID: 6W63). 

\subsection{\label{rosetta}Conformational ensemble generation}
We introduced backbone variability by generating 25 structures with the \textsc{backrub} application of the Rosetta protein modelling suite.\cite{kaufmann2010practically} These structures were further refined by repacking the side chains on these new backbone conformations with the \textsc{fixbb} application of the same package. These 25 structures make a conformational ensemble that span the timescale required to pivot the backbone along the alpha carbons and switch side chain rotamers, typically occurring in the microsecond timescale. 

The same protocol was followed after removing the complexed inhibitor from the starting PDB structure, obtaining 25 reference structures for the apo state of the SARS-CoV-2 main protease. 
\subsection{\label{tinker}Molecular dynamics with polarizable force field}
Each of the 25 starting structures in both the bound and apo states were solvated using the Gromacs \textsc{solvate} tool and a pre-equilibrated water box.\cite{pronk2013gromacs} Sodium and chloride ions were added to simulate physiological conditions while setting the overall charge of the system to zero. Each solvated structures was then minimized in Tinker\cite{rackers2018tinker} with the BFGS procedure and the AMOEBA polarizable force field.\cite{zhang2018amoeba,laury2015revised,ponder2010current} Parameters for the X77 inhibitor were derived from DFT generated electrostatic potentials, as described in Ref.\cite{ponder2010current} Molecular dynamics simulations were then performed in the NPT ensemble with a 1 fs timestep for 100 ps. The first 50 ps served as an equilibration step and the last 50 ps served as production data from which we can calculate electric fields. 
\subsection{\label{electric}Electric field calculations}
Electric fields were calculated every 1 ps of our production trajectories using our open-source code \textsc{ELECTRIC}.\cite{nash2020electric} An extensive documentation is provided in the Github repository where the code is hosted. The fields were projected onto the S-H bond of catalytic residue Cys-145 as well as the two C=O bonds of the amid groups of X77 (labelled C$_1$=O$_1$ and C$_2$=O$_2$). Averages were performed over the production runs (50 ps) as well as the 25 structures that make the conformational ensemble, thereby characterizing the electrostatic landscape of the enzyme across multiple time scales. 

\section{\label{S145}Enzyme dynamics and response to non-covalent binding}
\begin{figure}[h!]
\includegraphics{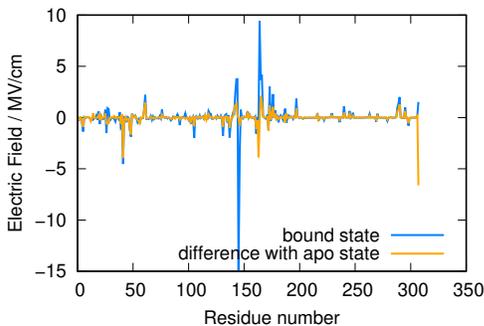}% Here is how to import EPS art
\caption{\label{fig:145} Electric fields projected onto the S-H bond of Cys-145. We present, in blue, the projected electric fields in the bound state, split into individual residue contributions. The last residue (\#307) is the solvent. We also show, in yellow, the projected electric field difference between the bound and apo state. A positive field supports breaking the S-H bond while a negative field promotes electron transfer from S to H.  }
\end{figure}

In this section, we look at the enzymatic response to the presence of X77 in the active site. We quantify this response by calculating the electric fields projected onto the S-H bond of the catalytic residue Cys-145. Figure \ref{fig:145} shows the contribution of each residue and solvent (residue \#307) to the projected electric field in the bound state. 

We observe notable contributions from residues His-41 (-4.5 MV/cm) , Cys-145 (-14.9 MV/cm), His-164 (9.4 MV/cm) and Glu-166 (4.2 MV/cm). Since the probe in these calculations is the S-H bond of Cys-145, we expect to see a large contribution from this residue. Further, the magnitude of this contribution does not change in the apo state, as illustrated by the projected electric field difference between the apo and bound states (Figure \ref{fig:145}). A similar observation could be made for Glu-166, leaving His-41 (the second catalytic residue), His-164 and water to govern the enzymatic response to X77 binding. 

\begin{figure}[h!]
\includegraphics{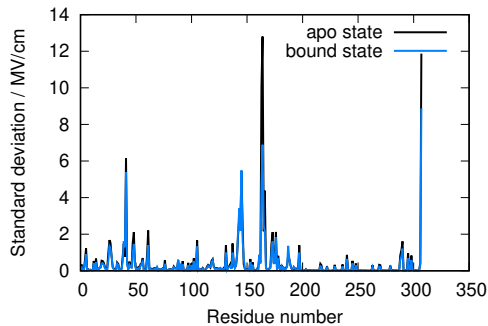}% Here is how to import EPS art
\caption{\label{fig:145_std} Standard deviation of the ensemble electric fields projected onto S-H bond of Cys-145. We show the standard deviation of each residue, including water (residue \#307) in the apo (black) and bound (blue) states.}
\end{figure}

\begin{figure}[h!]
\includegraphics[width=0.4\textwidth]{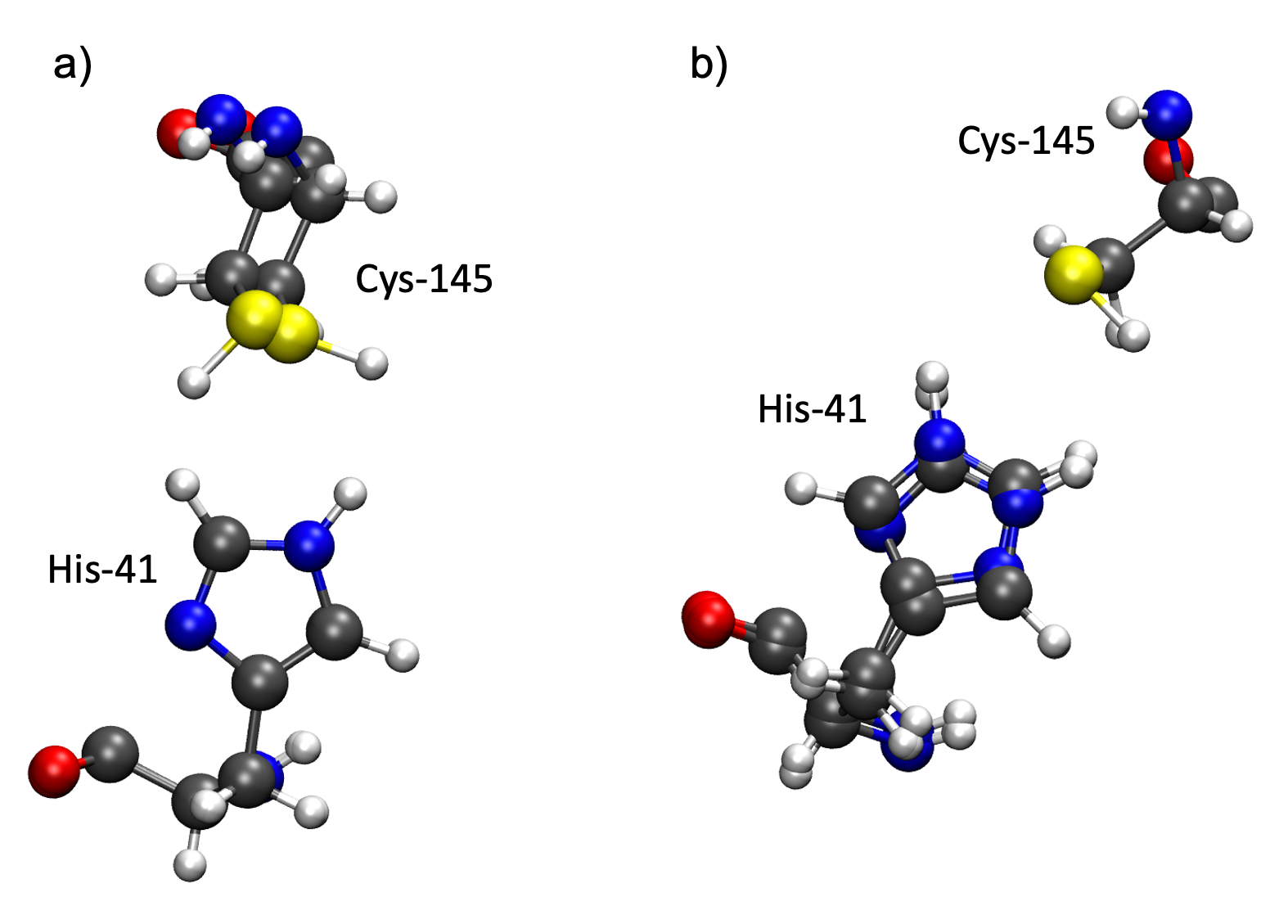}% Here is how to import EPS art
\caption{\label{alignement} Catalytic dyad conformations in the computed ensemble of SARS-CoV-2. (a) Change of rotamer in Cys-145 that reorients S-H bond towards His-41. (b) Conformation change in His-41.}
\end{figure}

Water contributes less to the projected electric field in the bound state because water molecules are being displaced out of the active site by the inhibitor. In the apo state, the electric field contribution from water is negative, which shows that the deprotonation of S-H is not favored. In the bound state however, the water contributes almost nothing to the projected electric field. This would suggest that the removal of active site water, rather than the specific chemical nature of the substrate, assists the ``activation" of the catalytic residue Cys-145 for nucleophilic attack. In contrast, His-41 contributes almost exclusively in the bound state, promoting electron transfer from S to H. This implies that the presence of the inhibitor in the active site modifies the portfolio of enzyme conformations that are representative of the ensemble in each state, selecting for conformations that results in a higher coupling of the two residues of the catalytic dyad. To better understand this effect, we analyze the values of the projected electric field average over each trajectory (tens of picosecond timescale) but not over the conformational ensemble (microsecond timescale). This allows us to look at the underlying heterogeneity of the ensemble and the corresponding conformation. This can be done by looking at the standard deviation of the electric field projected onto the S-H bond of Cys-145, as shown in Figure \ref{fig:145_std}. 
With a standard deviation reaching 6 MV/cm, we see that His-41 shows significant variability, confirming that multiple conformations are visited within the submicrosecond timescale. Looking at the 25 values that makes this average (see Appendix A), we note that His-41 contributes from -16.4 MV/cm to 11.9 MV/cm to the projected electric field. This means that the field average over the conformational ensemble hides the underlying heterogeneity of the ensemble, as previously observed for Ketosteroid Isomerase. The simulations with the highest average contribution from His-41 are primarily associated with conformations where the $\chi_3$ rotamer of the Cys-145 residue orients the S-H bond towards His-41 (Figure \ref{alignement}). Others are associated with a 180 degree rotation of the aromatic ring on His-41, as also shown in Figure \ref{alignement}. This alignment occurs more frequently when X77 is bound, which results in a higher average over the conformational ensemble.

The contribution of His-164 varies twice as much across the ensemble, with a standard deviation that reaches 12 MV/cm in the apo state. This suggests that His-164 is on a more flexible part of the protein and is subject to larger conformational changes than other residues. This flexibility is reduced in the bound state (standard deviation of 6 MV/cm) but the contribution to the projected electric field averages 0 across the ensemble (the residue that have functional relevance for the S-H bond of Cys-145 is the neighboring residues Met-165). This shows that the binding of X77 selects His-164 conformations that results in a stronger contribution of Met-165. In summary, the functional role of His-164 is binding the substrate, which we look at more closely in the next section.

\section{\label{Binding}Characterization of X77 binding}
\begin{figure}[h!]
\includegraphics{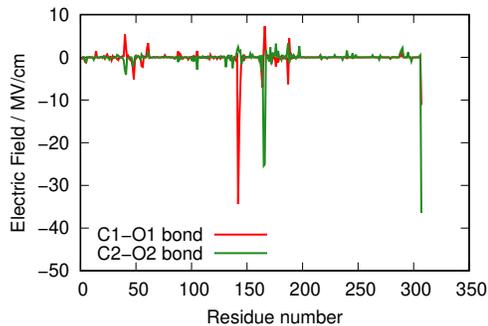}% Here is how to import EPS art
\caption{\label{fig:co} Electric fields projected onto the amid C$_1$=O$_1$ (red) and C$_2$=O$_2$ (green) bonds of X77. The total electric is shown split into individual residue contributions. A negative field supports breaking the double bond while a positive field promotes electron transfer from O to C. }
\end{figure}

In this section, we look at the contribution of each residue and the solvent to the electric fields projected onto the C=O bonds of X77 (C$_1$=O$_1$ and C$_2$=O$_2$ in Figure \ref{X77}). These two probes mimic the amid group in peptides that interacts with SARS-CoV-2 main protease (Cys-145 in particular), providing a direct handle on the intermolecular interactions between the inhibitor and the enzyme. Figure \ref{fig:co} shows the magnitude of the projected electric fields for both bonds. 

We observe that C$_1$=O$_1$ interacts predominantly with Asn-142, which exhibits a -34.4 MV/cm contribution to the projected electric field. In contrast, C$_2$=O$_2$ shows two identical contributions from Met-165 and Glu-166 (-25.4 and -25.0 MV/cm, respectively). This is the signature of two very localized interactions between the inhibitor and the enzyme, which does not involve the whole active site. More importantly, the two C=O bonds do not interact directly with Cys-145. Rather, X77 pins two residues on either side of the catalytic residue, blocking access to other molecules. Interestingly, while the residues that interact with C$_2$=O$_2$ have a functional impact on the S-H bond of Cys-145, Asn-142 seems exclusively involved in binding the inhibitor. We also note that water mediates the enzyme-inhibitor interaction, particularly via the C$_2$=O$_2$ bond. Water molecules in the active site therefore strengthen the binding of X77 to the enzyme. 

\begin{figure}[h!]
\includegraphics{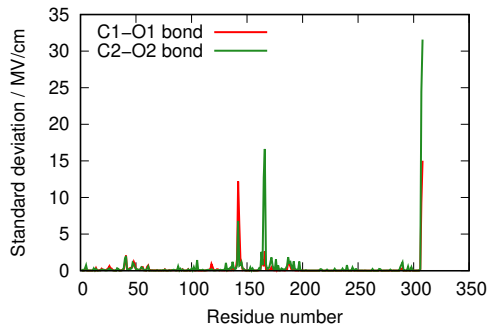}% Here is how to import EPS art
\caption{\label{fig:std} Standard deviation of the electric fields projected onto C$_1$=O$_1$ (red) and C$_2$=O$_2$ (green) bonds of X77. We show the standard deviation of each residue, including water (residue \#307).}
\end{figure}

The difference in behavior of C$_1$=O$_1$ and C$_2$=O$_2$ also manifests itself when looking at the standard deviation of the projected electric fields (Figure \ref{fig:std}). 

The standard deviation of the contribution of Asn-142 to the electric field projected onto C$_1$=O$_1$ is 13 MV/cm. This characterizes protein local conformation changes in the rotamer population of Asn-142 that directly impacts the binding of the inhibitor. However, while the standard deviation of the contribution of Met-165 and Glu-166  to the electric field projected onto C$_2$=O$_2$ is even higher (over 15 MV/cm), we also observe significant variability in the contribution of Asn-142 to the electric field projected onto C$_2$=O$_2$. This reveals a coupling between the two binding sites of X77 that can be utilize to strengthen the enzyme-inhibitor interaction. Finally, the standard deviation of water is high for both bonds (although especially for C$_2$=O$_2$), consistent with a higher number of possible water molecule orientations in the active site. 

\section{\label{conclusion}Conclusions}
\begin{figure}[h!]
\includegraphics[width=0.5\textwidth]{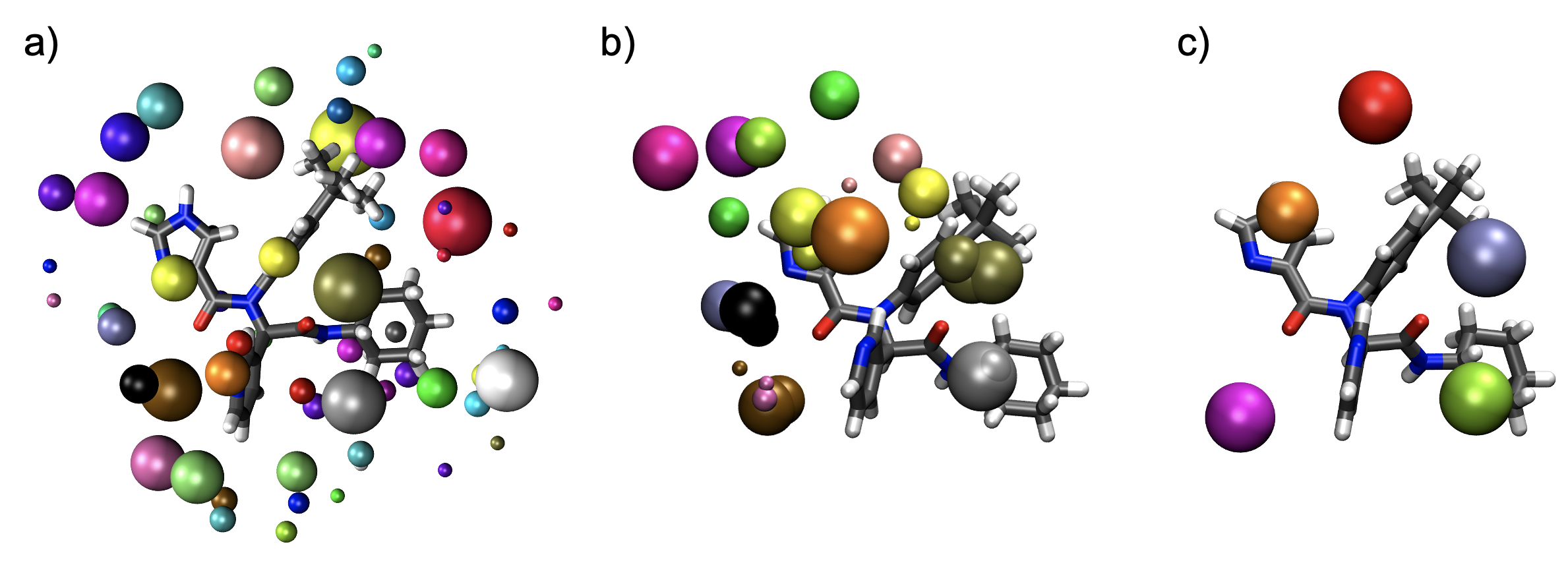}
\caption{\label{beads}Reduction of the search space using electric field calculations to identify the key residues involved in the binding of X77 in the active site of M$^\text{pro}$. (a) Selection of all residues within 5 \r{A} of the inhibitor, $\sim$40 residues . (b) Selection of all residues within 5 \r{A} of the binding elements of the inhibitor, namely the two C=O bonds of the amid groups, $\sim$15 residues (consistent with previous reports) (c) Selection of the residues that have significant impact on the projected electric fields of the system, 3 residues in addition to the catalytic dyad.}
\end{figure}

In summary, we reported in this paper the first MD-based electric field calculations used to understand enzyme-inhibitor interactions in the context of drug design. Electric field calculations can directly link structure to function in MD simulations, an ideal tool to identify the key residues involved in the binding of inhibitors in enzymatic active sites. We used a non-covalent inhibitor (X77) of the main protease of SARS-CoV-2 as an example given its relevance to the ongoing pandemic. Previous studies reported that over 15 residues could be involved in substrate binding to the protease, considerably complicating the search for new inhibitors. Using electric field calculations, we found that X77 binds the active site via two very localized interactions, involving only a couple of residues. The binding sites allow the inhibitor to block the catalytic residue Cys-145 from other molecules, preventing further enzymatic activity. More specifically, we found that Asn-142 and Met-165 (in cooperation with Glu-166) modulates the interaction of X77 with M$^\text{pro}$. This is a significant refinement from our previous understanding of M$^\text{pro}$, as illustrated in Figure \ref{beads}. We also found that water plays a significant role in mediating X77-M$^\text{pro}$ interactions via the C=O bonds. The findings in this paper can be used to narrow the exploration space when optimizing close contacts in the search for non-covalent inhibitors, many of which have already been made to mimic X77. Overall, our work has the potential to accelerate the route towards the production of new SARS-CoV-2 remedies, and can be applied to other enzyme inhibitors. 

\begin{acknowledgments}
The author thanks the Virginia Tech Department Faculty Start-up Funds for financial support and acknowledge Advanced Research Computing at Virginia Tech for providing computational resources and technical support that have contributed to the results reported within this paper.
\end{acknowledgments}

The data that support the findings of this study are available from the corresponding author upon reasonable request.

\appendix

\section{Appendixes}

\begin{table}[h!]
\caption{\label{tab:ensemble}This is a narrow table which fits into a
text column when using \texttt{twocolumn} formatting. Note that
REV\TeX4 adjusts the intercolumn spacing so that the table fills the
entire width of the column. Table captions are numbered
automatically. This table illustrates left-aligned, centered, and
right-aligned columns.  }
\begin{ruledtabular}
\begin{tabular}{cccccc}
Simulation &His-41 &His-164 &Asn-142 &Met-165 &Glu-166\\
 &Cys-145 &Cys-145 & C1=O1 &C2=O2&C2=O2\\
\hline
1 & -4.5&15.6&-29.3&-23.6&-26.8\\
2 & -6.3&6.7&-48.0&-33.6&-32.4\\
3 & -2.1&14.3&-32.8&-32.7&-32.5\\
4 & -11.4&19.3&-48.6&-23.0&-26.8\\
6 & -7.9&11.9&-18.9&-22.3&-28.8\\
7 & -2.3&3.4&-31.1&-24.2&-28.2\\
8 & -7.1&4.9&-32.5&-33.8&-32.5\\
9 & -1.1&9.3&-36.2&-29.0&-5.5\\
10 & -5.8&13.2&-23.6&-29.9&-30.3\\
11 & -8.7&14.6&-30.4&-28.4&-38.0\\
12 & -7.2&13.4&-34.4&-30.3&-29.4\\
13 & -0.9&2.8&-49.0&-29.1&-33.7\\
14 & -0.7&12.6&-19.7&-21.0&-25.8\\
15&-7.5&5.1&-16.2&-20.0&-22.3\\
16&-4.5&21.5&-47.2&-30.8&-27.4\\
17&-16.4&19.8&-41.4&-29.7&-41.2\\
18&0.2&3.1&-35.4&-25.4&-31.6\\
19&-1.5&4.1&-38.2&-36.1&-30.3\\
20&-8.6&5.7&-34.7&-12.8&8.1\\
21&-7.0&4.5&-47.1&-30.3&-35.7\\
22&-6.5&13.1&-33.5&-25.7&35.9\\
23&-7.4&13.9&-13.0&-31.4&-20.3\\
24&2.9&-0.2&-44.4&-33.9&-32.5\\
25&11.9&-6.6&-38.7&-25.5&-32.3\\
\end{tabular}
\end{ruledtabular}
\end{table}

\bibliography{main}

\end{document}